\documentclass[fleqn,10pt]{wlpeerj}

\usepackage{inconsolata}
\usepackage{microtype}
\usepackage{url}
\usepackage{xcolor}
\usepackage{hyperref}
\usepackage{cleveref}

\hypersetup{
    colorlinks,
    linkcolor={magenta},
    citecolor={cyan},
    urlcolor={magenta}
}

\title{A Mixed-method Study on Security and Privacy Practices in Danish Companies}

\author[1]{Asmita Dalela}
\author[2]{Saverio Giallorenzo}
\author[1]{Oksana Kulyk}
\author[3]{Jacopo Mauro}
\author[1]{Elda Paja}

\affil[1]{IT University of Copenhagen, Denmark \texttt{\{ asmd, okku, elpa \}@itu.dk}}
\affil[2]{Universit\`a di Bologna, Italy and INRIA, France \texttt{saverio.giallorenzo@gmail.com}}
\affil[3]{University of Southern Denmark, Denmark \texttt{mauro.jacopo@gmail.com}}

\begin{abstract}
Increased levels of digitalization in society expose companies to new security threats, requiring them to establish adequate security and privacy measures. Additionally, the presence of exogenous forces like new regulations, e.g., GDPR and the global COVID-19 pandemic, pose new challenges for companies that should preserve an adequate level of security while having to adapt to change. In this paper, we investigate such challenges through a two-phase study in companies located in Denmark---a country characterized by a high level of digitalization and trust---focusing on software development and tech-related companies. Our results show a number of issues, most notably i) a misalignment between software developers and management when it comes to the implementation of security and privacy measures, ii) difficulties in adapting company practices in light of implementing GDPR compliance, and iii) different views on the need to adapt security measures to cope with the COVID-19 pandemic.
\end{abstract}

\begin{document}

\flushbottom
\maketitle
\thispagestyle{empty}

\section{Introduction}
\label{sec:introduction}

The fact that security and privacy are a challenge to companies has long been accepted in research, requiring both technical solutions and a consideration of human and societal factors~\cite{G12}. Moreover, the ever-growing presence of digital services in people's everyday life and the evolving landscape of security and privacy threats, require companies to adapt to new challenges to avoid severe consequences such as data theft or loss of reputation \cite{equifax}.

Previous studies have shown that the proper implementation of security and privacy processes in companies is often lacking, even for companies employing people with a high level of technical expertise such as software development companies~\cite{balebako2014privacy,weir2020needs,van2019data,haney2018we,assal2018security}. Security and privacy processes as a subject in need of continuous change and adaptation, however, are less documented.

In this work, we investigate the challenges faced by Danish companies in implementing and keeping up to date security and privacy measures. 
Since Denmark is a highly digitalized country, it has a high dependency on secure digital solutions that call for a high degree of data protection practices, as well as adequate security measures to be adopted by Danish companies. In our study, we focus on software development and tech-related companies, investigating how they deal with security and privacy: (1) when conducting their day-to-day practices, (2) when required to adapt to new legislation, namely, the introduction to the EU General Data Protection Regulation (GDPR), and (3) when required to adapt to an unforeseen crisis during a rapidly and unpredictably changing situation, namely, the COVID-19 pandemic and the introduced restrictive measures. Our contribution addresses the following research objectives:

\begin{itemize}
    \item In terms of organizational practices, how are security and privacy integrated? How are the responsibilities defined and what are the controls (if any) that are implemented? How do companies ensure sufficient security and privacy competences among their employees?
    \item In light of GDPR entering into force, how have the companies been dealing with it? Do they incorporate the required measures towards compliance, and what are the challenges they are facing in doing so?
    \item In light of the COVID-19 pandemic, how have companies adapted to the situation? What are their concerns and challenges given the need to shift to remote work?
\end{itemize}

Our investigation identifies nuances suggesting that companies lack proper guidelines to support them in adapting to a diversity of emerging challenges. These nuances include
a lack of knowledge of proper security and privacy measures and a lack of awareness about security and privacy risks, leading to a situation where the necessary changes such as GDPR compliance measures and remote-work policies are not being fully implemented. 

Furthermore, the results show that there is a misalignment between the perception of security issues and responsibilities of senior management, software developers, and people responsible for security and privacy in the company. This presents a barrier to ensuring that the employees have the necessary competences for implementing the security and privacy measures and that they are given a proper opportunity to do so.

We also note the occurrence of trust, in terms of social cohesion\footnote{Defined by Larsen~\cite{L13} as ``the belief that [citizens] share a moral community, which enables them to trust each other''.}, as a common underlying thread across different themes. While the role of trust in determining security measures is often stressed by previous research in other contexts \cite{haney2018,ashenden2013cisos}, trust is considered to be a distinctive cultural value in Denmark and other Nordic countries~\cite{L13,sonderskov2014danish}. Such importance of trust, as a distinguishing characteristic of the Danish society, has been specifically noted in our study, allowing us to elaborate on both positive and negative effects of it on security and privacy.

\section{Related Work}
\label{sec:related}
Previous research on security and privacy challenges in organizations revealed that a source of problems is the difficulty of the employees to comply with the security policies~\cite{beautement2008compliance,mayer2017reliable,das2014,blythe2015,ashenden2013cisos,haney2018}. In particular, these studies have identified several behavioral factors, which influence compliance to security and privacy policies, including the perceived severity of threats, self-efficacy, trust between the employees and the security team, perceived usefulness of the policies, costs of following the policies, the severity of sanctions for non-compliance or social influence perceived norms in one's environment.

In particular, the need for addressing the human factors of security in software development has been highlighted in recent years, e.g., with Green and Smith~\cite{green2016} indicating
developers as the ``weakest link'' and Acar et al. proposing a research agenda for such investigations~\cite{acar2016you}. Specifically, studies have been conducted to study different aspects of software development, such as the adoption and usability of specific tools (e.g., static analysis tools~\cite{smith2020can} and cryptographic APIs~\cite{acar2017comparing}), available guidance and support materials~\cite{acar2017developers}, organizational processes in software development companies~\cite{haney2018we,assal2018security,PTLLO20}, and individual behavior and mental models of security and privacy of software developers~\cite{balebako2014privacy,weir2020needs,van2019data,xiao2014}. Many of these works have been summarized in a systematic literature review by Tahaei and Vaniea~\cite{tahaei2019survey}. Overall, these studies reveal a variety of issues, such as the complexity of existing tools and procedures, the lack of security-focused expertise among developers, the lack of reliable guidance, and the prioritization of functional features over security, altogether stressing the importance of establishing a security culture within the company.

Another research strand has investigated the cultural aspects of security and privacy. As such, studies of leaked passwords from different countries (India, Japan and the UK~\cite{mori2020comparative}, as well as the US and Germany~\cite{mayer2017second}) reveal the differences in password complexity and chosen words, the most common being culture specific. 
Other studies have shown the differences in security and privacy risk awareness and behavior comparing for instance participants from Spain, Romania and Germany~\cite{kulyk2020security} and participants from Germany, the UK and the US~\cite{coopamootoo2020dis}. Among the studies on these topics, ~\cite{volkamer2018replication} is particularly relevant for our work, as it involves another Nordic country. In this work, Volkamer et al. investigate the differences in taking security precautions during ATM usage (e.g., whether people hide their PINs during cash withdrawal) among the participants from Germany, UK and Sweden. The study shows that the participants from Sweden and the UK were less likely to take precautions, suggesting the difference in cultural norms as the reason for these differences. Designing security awareness and education measures in different cultural contexts has been studied by Bada et al.~\cite{bada2019cyber}, comparing the security awareness campaigns in the UK and Africa and by Al Qahtani et al.~\cite{al2018effectiveness}, replicating the US study on the effectiveness of an awareness campaign in Saudi Arabia and adjusting the campaign contents towards the Saudi cultural context. Both of these studies show that cultural characteristics, such as shared values (e.g., individualism vs. social responsibility) or specific threats that are prevalent in a specific society, are an important factor in shaping these campaigns.

Following the findings from previous research, we look at security as a social issue, considering the dynamics between people in different roles in organizations and the interconnection of the perspectives they have on security and privacy issues, as well as the influences of a broader culture. Our study explores these perspectives in the context of Danish companies, taking into account the high level of trust and digitalization in the Danish society and the recent challenges the companies have been confronted with.

\section{Methodology}
\label{sec:methodology}
\newcommand{\circled}[1]{\raisebox{.5pt}{\textcircled{\footnotesize\textsf{#1}}}}

We dedicate this section to present the mixed-method approach
used in this work, detailing its parts: a quantitative
survey (\Cref{sec:survey}) and a subsequent round of interviews
(\Cref{sec:interviews}). We conclude the section with a discussion on the ethical considerations of our studies (\Cref{sec:ethics}).

Our mixed-method approach follows a two-phase \emph{sequential explanatory design}~\cite{ICS06}.
 \begin{figure}[htb]
 	\centering
 	    \centerline{\includegraphics[width=0.49\textwidth]{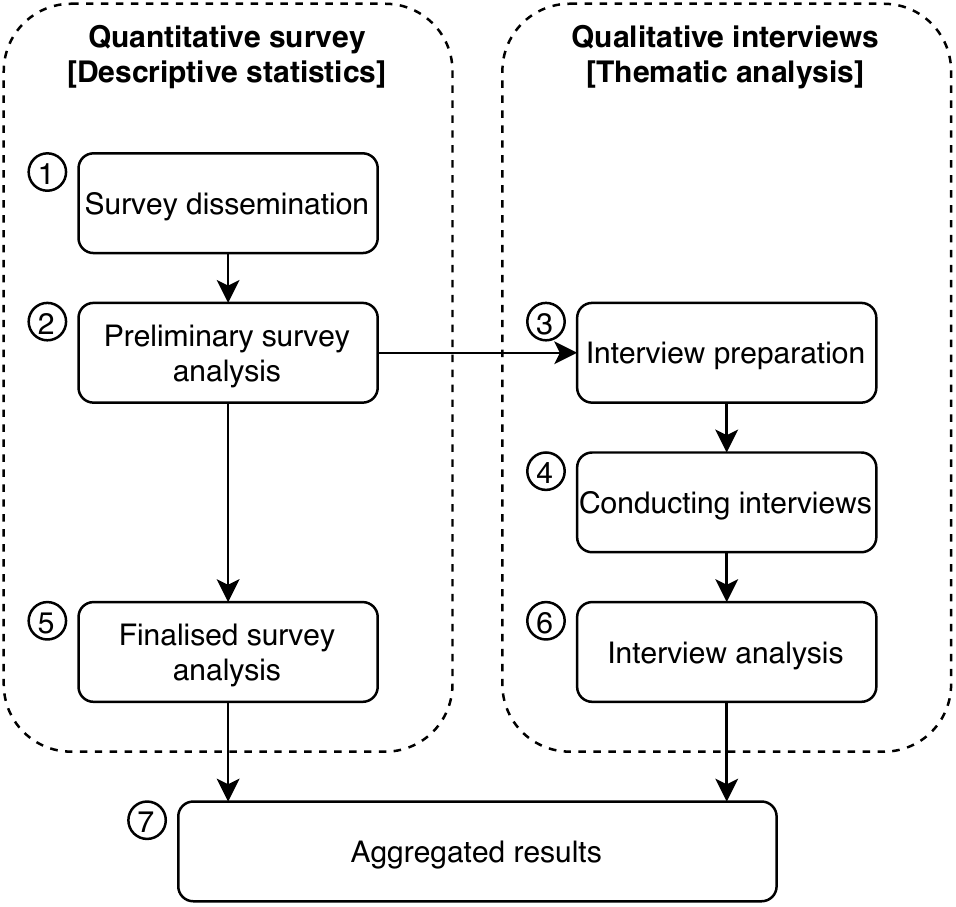}}
 	    \caption{Scheme of the methodology followed in this
 	    study.\label{fig:mixed-meth}}
 \end{figure} 
%
As visualized in \Cref{fig:mixed-meth}, we conducted a survey to
collect data for a first quantitative evaluation \circled{1}, and used the preliminary results collected from the survey \circled{2} to inform the interview preparation, namely, the development of the interview guide~\circled{3}. Then, we conducted ethnographic interviews \circled{4} to gain in-depth qualitative insights over the selected aspects. 
As the data collection period for the survey overlapped with conducting the interviews, we finalized the analysis of the survey \circled{5} together with the analysis of the interviews \circled{6}, aggregating findings from both the quantitative and the qualitative part \circled{7}. 

\subsection{Quantitative Survey}

\label{sec:survey}
The first stage of the study was done as an online survey. The goal of the survey was to obtain initial quantitative insights into security and privacy practices in companies across five areas: i) security management and standards, including challenges in adhering to these standards; ii) the integration of security into the development cycle; iii) the integration of GDPR; iv) general perception of security awareness, security policies, behavior, reporting and available training; v) the impact of pandemic on security. To get a broad perspective on these areas, the survey aimed at eliciting responses from respondents occupying different roles (e.g., management, developers). We describe the survey design, dissemination, and the resulting sample in more detail below.


\subsubsection{Survey design}
The survey consisted of a total of $36$ questions, divided across the five areas that were the focus of the investigation. At the beginning of the survey, the participants were asked about their roles in the company, and were encouraged to choose from a predefined list of tasks,
viz. management related, IT-security related, privacy/data-protection related, software-development related, IT-administration related and `other'. The participants in the survey had the option to choose multiple organizational roles, if they considered that the combination of the given roles better described their position.

The respondents were then shown the questions tailored to the roles they selected so that only 8 out of the 36 questions were presented for each role. Still, the participants were allowed to skip any question they did not want to answer. The full survey questionnaire can be found in \Cref{sec:survey_questionnaire}. 

Before launching the survey, two runs of pretest were conducted. 
The pre-testers included experts in cybersecurity, human-factors research, and software development, from both academia and industry, and all residing outside Denmark. 
The pretests validated different aspects of the survey, such as the clarity of the questions, its duration, and all the possible role selections with the corresponding questions. The required time to complete the survey was determined to be 10 minutes.


\subsubsection{Survey dissemination}
The survey was implemented as a questionnaire hosted on the SurveyXact platform\footnote{\url{https://www.survey-xact.dk}, last visited on February 2021}
 based in Denmark.
These companies were from diverse sectors such as software product development, pharmaceuticals, retail, manufacturing, finance. The companies were categorized into two groups: small and medium enterprises (SMEs) ($\leq$ 250 employees) and large ($>$ 250 employees). For the second dimension, eight relevant participant roles were identified to send the survey to: CEO, CTO, CISO, DPO, developers, IT administrators, HR, and finance. Irrespective of the size of the company, the survey was sent to its CEO,
requesting to disseminate it to the other relevant participants.\footnote{Note that CEO, CTO, CISO, DPO and HR are acronyms for Chief Executive Officer, Chief Technology Officer, Chief Information Security Officer, Data Protection Officer and Human Resource manager respectively.}

The survey ran from June to November 2020, 
and it was promoted in two phases:
first in mid-June, and again in early August, to maximize its
reach within companies.\footnote{Note that July is the holiday month in Denmark and thus we did not do any promotional activity during this month.} To maximize the reach-out to the relevant participant roles, five different channels were leveraged for the survey promotion: social media, trade bodies, startup accelerators, the internal network of the authors' universities, and media publications.

\subsubsection{Survey sample and analysis}

\begin{table}[t]
\centering
\begin{tabular}{||c|c|c||}
\hline
Role & SMEs & Large companies \\ \hline \hline
Management & 38 & 14 \\ \hline
IT-security & 24 & 19 \\ \hline
Privacy/data protection & 15 & 8 \\ \hline
Software development & 18 & 18 \\ \hline
IT administrator & 21 & 11 \\ \hline
Other & 6 & 11 \\ \hline
\end{tabular}
\caption{Number of participant selections for each role (participants could
select multiple roles).\label{tbl:roles}}
\end{table}

Overall, 107 participants completed our survey, of them 47 from large companies and 60 from SMEs. \Cref{tbl:roles} shows the distribution of the participants' roles in the companies. The analysis was done in an exploratory way, preparing the descriptive statistics related to our research objectives and serving as the groundwork for the next study phase.

\subsection{Qualitative Interviews}

\label{sec:interviews}

In this section, we describe how the interviews were planned, their reach-out strategy, and conclude with the methodology used for their analysis.

\subsubsection{Interview structure planning}

The initial insights from the survey were used as a basis to discover the main areas for in-depth investigation during the ethnographic interviews. These interviews consist of a conversation between a researcher (interviewer) and interviewee, where knowledge is constructed in the interaction between them~\cite{S16}.

Interviews took place from September to November 2020, following a preliminary analysis of the survey conducted in August 2020. 
To adapt to the COVID-19 containment regulation, most interviews were conducted over video calls using Microsoft Teams.

Interviews were planned by creating an Interview Guide (\Cref{sec:interview_guide}) with inter-related questions aimed at drawing-out the perspective of the interviewee. They were conducted through semi-structured conversations lasting 1 hour.

\subsubsection{Reach-out strategy and interviewee recruitment}
We decided to conduct interviews with specific participants chosen across two dimensions: participant role and company size.
For the participant role dimension, we aimed at covering different points-of-view of the interviewees on the same issues such as general security and privacy integration, the effect of GDPR, impact of the pandemic on security implementation and others, for triangulating the perspectives, avoiding anecdotal conclusions, and drawing nuanced insights. Three participant roles were covered: senior managers, security experts and people responsible for security and privacy policies in the company, and developers. For the company size dimension, they were segmented into two broad  categories: SMEs and large companies 

The potential participants for the interviews were identified by leveraging different channels such as university networks, LinkedIn, Google, and reaching out and engaging through emails and LinkedIn messages.
The profile of the 11 participants can be viewed in \Cref{tbl:interviewees}.
The interviews were transcribed and anonymized to keep the opinions and identities of the participants secured. 

\begin{table}
\centering
\begin{tabular}{||c l c l||} 
 \hline
 \# & Role & Org Size & Sector \\ 
 \hline\hline
 1 & Senior Manager & SME & Software products \\ 
 \hline
 2 & Sec/Priv Expert & SME & Software products \\
 \hline
 3 & Senior Manager & SME & Software products \\
 \hline
 4 & Developer & SME & Software products \\
 \hline
 5 & Senior Manager & SME & Finance \\
 \hline
 6 & Developer & SME & Construction \\
 \hline
 7 & Developer & Large & Services \\
 \hline
 8 & Senior Manager & Large & Services \\
 \hline
 9 & Sec/Priv Expert & Large & Retail/CPG \\
 \hline
 10 & Sec/Priv Expert & Large & Manufacturing \\
 \hline
 11 & Developer & Large & Manufacturing \\ 
 \hline
\end{tabular}
\caption{Profiles of the Interview participants.\label{tbl:interviewees}}
\end{table}

\subsubsection{Interview data analysis}
The analysis of individual interviews was conducted using the thematic analysis methodology~\cite{BC06} to distill a set of key themes across all the ethnographic interviews. Following this methodology, chosen themes were selected when representing
some level of patterned response or meaning within the data set and capturing something important about the data. 

We took an inductive, open approach while establishing the themes based on the frequently appearing responses rather than aligning the participants' opinion to the preassigned categories. Following Michalec et al.~\cite{ODSA20}, we iteratively discussed the transcripts to construct the emerging themes and
fostered the discussions between the researchers (authors) to build a shared understanding.

\subsection{Ethical considerations}
\label{sec:ethics}
While our institutions do not have a mandatory Institutional Review Board for studies, we addressed  the four considerations related to ethics in such a research, namely, informed consent, confidentiality, consequences and the role of the researcher \cite{delamont2018,brinkmann2017}.

While conducting our study, we ensured to apply the ethical principle of confidentiality, so that private data identifying the participants will not be reported. 
As such, we also followed a set of guidelines when conducting this study, in line with the General Data Protection Regulation (GDPR). 

We explicitly obtained a consent from the interviewees, ensured a voluntary participation of the people involved, and informed them of their right to withdraw from the study anytime. Before starting the interviews, we clearly stated the goal of our study and data handling procedures. 

We also followed the ethical principal of beneficence where we looked that the knowledge gained through conducting this research should outweigh the risk of harm to the participant. Also, the role of researcher was critically reported to the participants at the beginning of our study. 


While we asked for some data that might be seen as personal during the survey (e.g., through a  number of the questions related to the role, organization-sector, organization-size, contact email for possible follow-up interviews), we explicitly made it clear on the start page of the survey that these questions (with an exception of a question asking about one's role) were not mandatory to answer, and any information that could be used to identify them would remain confidential. In addition, participants had the option to discontinue the survey at any time, while we emphasized that they could skip questions they preferred not to answer to for whatever reason. Finally, we did not provide any remuneration for our participants.

Since the interviews were conducted during the COVID-19 pandemic, there was a clear recommendation from the government to maintain social distancing and restrict traveling. We provided a choice to our interview participants to conduct the interviews on-premise or over video through MS Teams.

\section{Results}
\label{sec:results}
In this section, we discuss the results of both the survey and the interviews.
We focus on three key themes that emerged from the analysis of the ethnographic interviews and we assimilate them with the findings from the survey: i) general security and privacy integration, ii) effects of
the GDPR, and iii) effects of the COVID-19 pandemic\footnote{The results on the
other key topics are detailed in the full technical report (reference omitted for double-blind reviewing).}.


\subsection{General security and privacy integration}

\begin{figure}[htb]
	\centering
	    \centerline{\includegraphics[width=0.8\textwidth]{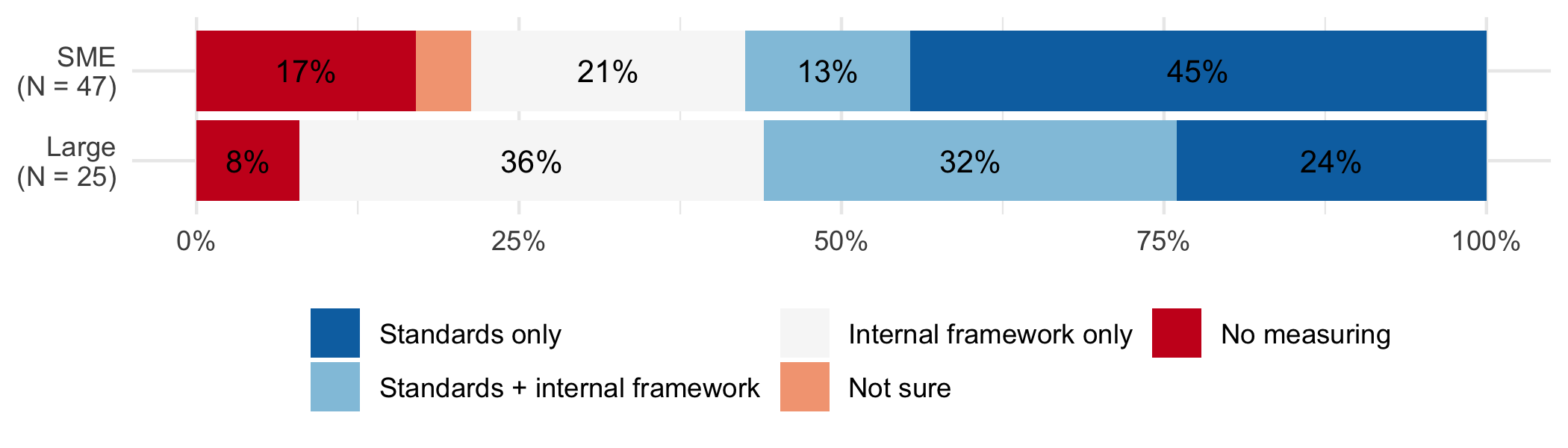}}
	    \caption{Approaches for measuring security
	    readiness.\label{fig:sec_measuring}}
 \end{figure}

Security professionals and managers (47 respondents from SMEs and 25 from large companies\footnote{Here and everywhere else in this section, we provide the number of participants who chose to answer the question. Note that since the participants could skip any question, the total number of responses for each question can differ.}) 
were asked about the general approaches the company takes in measuring cybersecurity readiness. 
As \Cref{fig:sec_measuring} shows, more than half of them ($58\%$ and $56\%$ of respondents in large companies and SMEs, respectively) reported relying on established standards either fully or in combination with frameworks developed internally in the company. 
A relatively small percentage (17\% and 8\% of respondents in SMEs and large companies respectively) reported not using any kind of measuring approaches at all. 
Furthermore, respondents from large companies were more likely to report on using internal frameworks, either as the only tool or in combination with established standards 
($68\%$ compared to $34\%$ respondents from SMEs).

\begin{figure}[htb]
	\centering
	    \centerline{\includegraphics[width=0.8\textwidth]{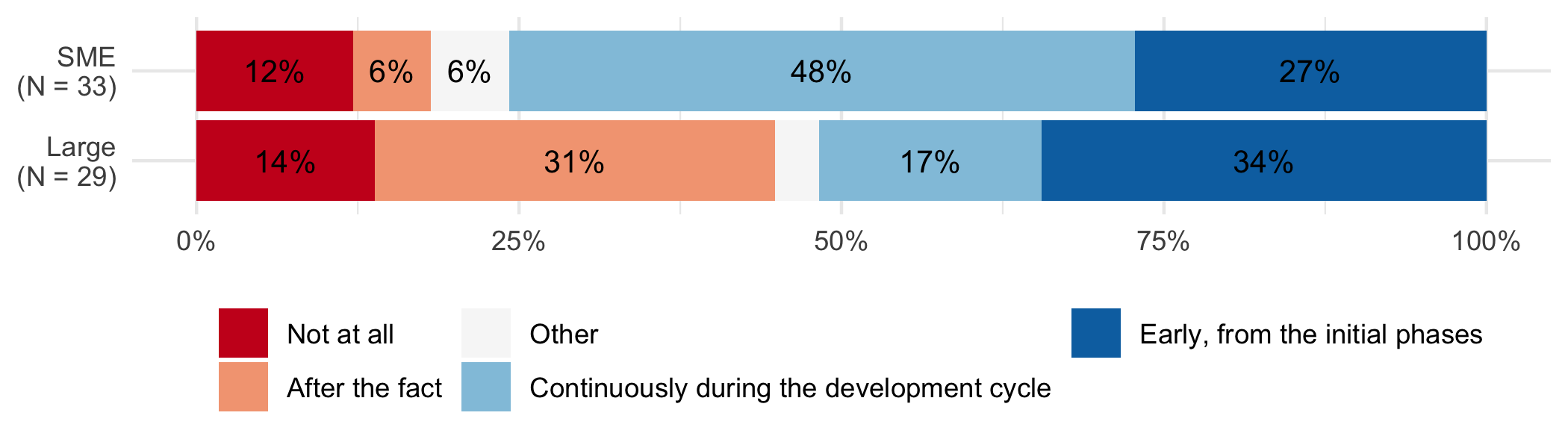}}
	    \caption{Integration of security into the development
	    cycle.\label{fig:dev_integration}}
 \end{figure}

The software developers who participated in the survey (33 respondents from SMEs and 29 from large companies) were asked about the stage in which security is integrated into the software development cycle. 
As shown in
\Cref{fig:dev_integration}, the majority of the respondents (75\% of respondents from SMEs and 51\% from large companies) reported security integration either early from the start or continuously during the development. 
However, almost half of the respondents from large companies (45\%) reported integrating security either after the fact, or not at all, in contrast to only 18\% of respondents from SMEs.

When asked about experience with security training (respondents in all roles, overall 50 from SMEs and 42 from large companies), 
the majority (56\% in SMEs and 76\% in large companies) reported either participating in or at least being aware of such training in their companies. 
At the same time, only 50\% of the respondents from SMEs participated in such training, and while this percentage was higher in large companies (69\%), many of them (24\%) did not find the training they attended to be value-adding.

 \begin{figure}[htb]
	\centering
	    \centerline{\includegraphics[width=0.8\textwidth]{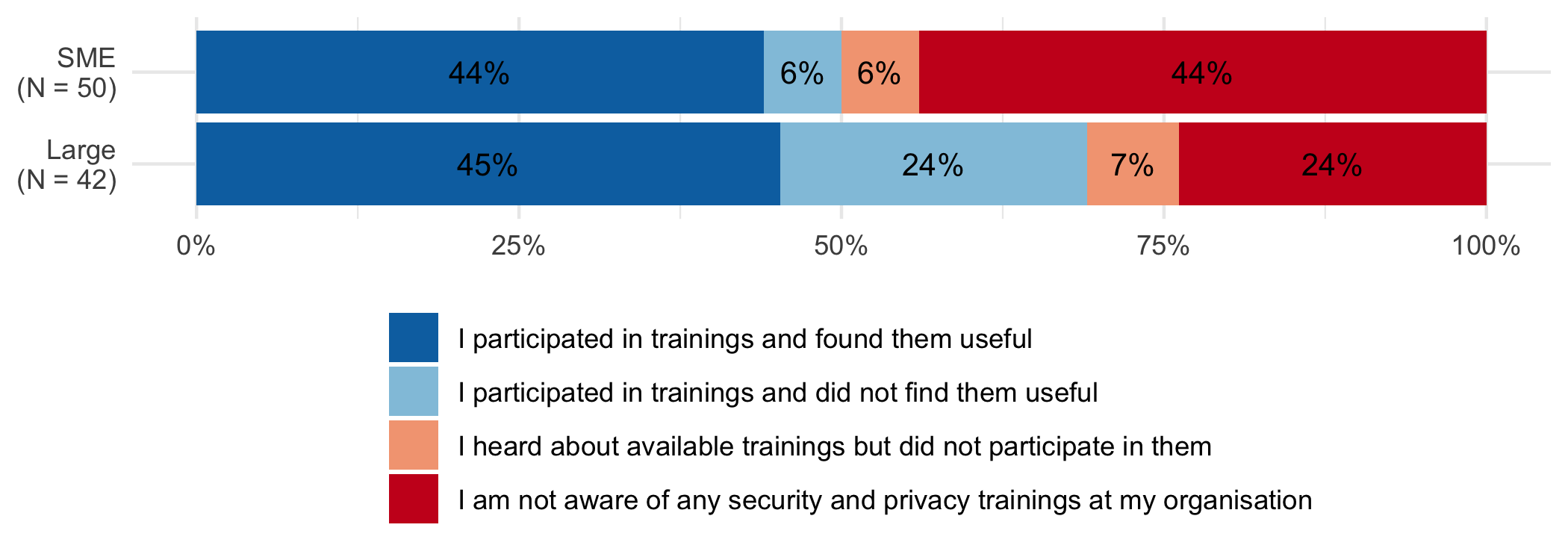}}
	    \caption{Training attendance and utility perception.\label{fig:sec_trainings}}
 \end{figure}

The analysis of the interviews revealed multiple nuances
potentially affecting the organizational practice of security and privacy measures integration in the development cycle, namely, the nuances of responsibility, trust, prioritization and competences. 

\paragraph{Responsibility.}
A common thread in our observations was the delegation of the responsibility to the developers, 
with senior management assuming that developers know and do all that needs to be done for implementing security for the product development. A senior manager articulated: 

\begin{quote}
``They [software developers] think about it [security] all the time and they are even 10 times more security-aware than I am.''
\end{quote}

Such a perspective was reinforced by some of the developers, confirming that the task of implementing security and privacy measures is often placed entirely on them:

\begin{quote}
``[Security and privacy measures] is not something that they [senior management] really encourage and it's not something they really care about. It's something that happens only if the engineers [developers] themselves take care of it.''
\end{quote}



\paragraph{Trust.} Trust was often mentioned in different contexts, as senior managers trust the developers in their company to have all the necessary knowledge and capabilities to take proper care of security and privacy in the development cycle. A senior manager remarked about the high level of trust in his developers.
%
%

\begin{quote}
 ``On the development side, the people who are working are extremely security-aware
 [...] I would say, I trust these people. I actually trust them a lot.''
\end{quote}

Another emerging context was the fact that companies tend to trust their employees to not intentionally engaging in malicious actions towards the company.
In particular, one interviewee commented that such prevalence of trust is a cultural characteristic of the Danish society:

\begin{quote}
    ``And I think Denmark as a national culture seems to be very trusting.''
\end{quote}

A complementing perspective on trust was also visible with a few senior managers who emphasized that the developers can trust their company enough to come forward with the reporting of security incidents.


\begin{quote}
``We're not a company where people are being shot in front of the building, if they come forward and say, `look, we think we have a problem [security breach] here'. So I'm hoping people will come forward if they do find a breach.''
\end{quote}


\paragraph{Prioritization.}
A common perspective among developers was that there was a lack of security prioritization on behalf of senior management.
The developers perceive that senior management often focuses more on the roll-out of functionalities, and they do not proactively and meaningfully prioritize the security needs as a part of their business imperatives. 

This approach was leading to a tug-of-war between functionality vs security mindset in the development teams. Some developers mentioned that their companies focus more on delivering functionality rather than security as it will decrease their time to market.

\begin{quote}
``So stuff like security, management didn't really want to spend time on or hear about, because that would delay whatever things we were supposed to deliver.''
\end{quote}

This often results in developers first working on the business requirements which deliver `something', and apply the security measures on their own, later, resulting in siloed implementations, 
and a superficial complacency about security, across the company.

Furthermore, the lack of involvement from the senior management was perceived as an issue in allocating resources towards cybersecurity. A developer highlighted

\begin{quote}
``I don't get the feeling that management, in general, know a lot about security or really invests time and resources into making this an important thing in our daily engineering.''
\end{quote}




\paragraph{Competencies.}
When it comes to acquiring competencies necessary for implementing security measures optimally, many interviewees mentioned that there is neither provisioning of general security awareness training nor any developer-specific training, in their companies.

\begin{quote}
``And not in this company or the other[company], there was any kind of mentions of security as part of the on-boarding. Then there are no courses or training or anything afterwards.''
\end{quote}

Even in cases when such training was available outside the company, some interviewees felt that participation in these training is generally discouraged and the senior managers want a justification for attending them:

\begin{quote}
``I was once or twice to some compliance and stuff [security training], but my management was not so pleased about, and you spend the time on things like these is such a big thing.''
\end{quote}

This is also reflected in the commentary from some senior managers, wherein they explained their view of generally encouraging developers `if' multiple criteria were satisfied from their perspective. One senior manager commented:

\begin{quote}
``Generally, yes we would encourage that [providing training] if people came forward, but it has to make sense, you know, where, what is their position, what is the purpose, what is the value to the company if they were to do it.''
\end{quote}





Only a few interviewees highlighted that their companies have a thorough approach to security training, including plans for specific security training for the developers to increase their proficiency level so that they can embed security in the development cycle. 
One of the security experts remarked:

\begin{quote}
	``My ideal scenario, we get a training program that says, here's the common body of knowledge for agile developers across all of these domains that we support. Here are some training modules, here's a platform that you can develop on and that will give you coaching as you go, to help you get better at writing secure code. And we want to start a network of champions within the agile squads.''
\end{quote}

\subsection{Effects of GDPR}
\begin{figure}[htb]
	\centering
	    \centerline{\includegraphics[width=0.8\textwidth]{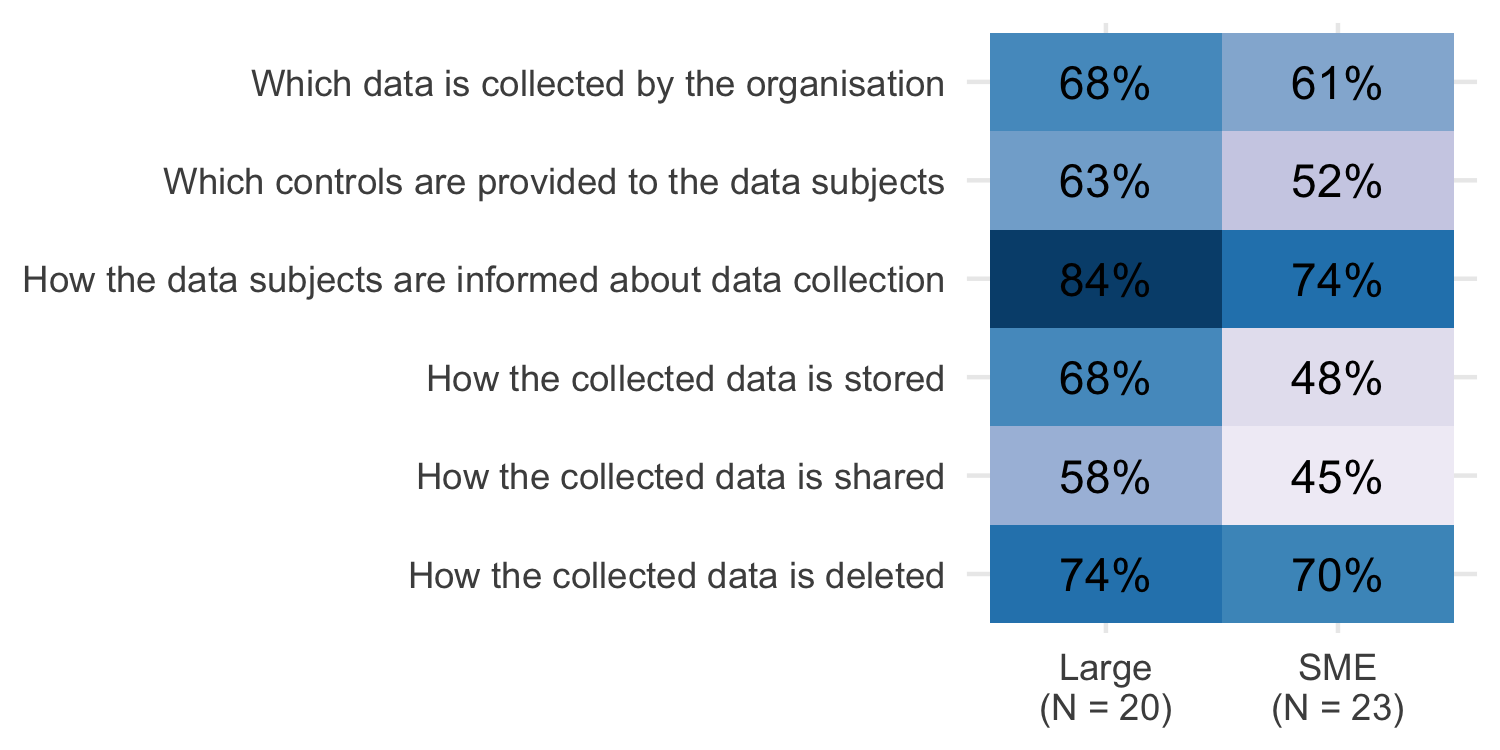}}
	    \caption{Percentage of respondents reporting changes in how the
	    company handles personal data since the GDPR entry into
	    force.\label{fig:gdpr_changes_yes}}
 \end{figure}

The survey participants who reported being responsible for either security or privacy-related tasks (43 in total) were asked about changes in their company since the GDPR entering into force. \Cref{fig:gdpr_changes_yes} shows the percentage of respondents who reported changing some aspects of data sharing,
namely, which data is collected, what controls are provided to the data subjects, how the data subjects are informed about the data collection, 
how the collected data is stored, shared and deleted. The results show that, overall, large companies were more likely to enact changes, 
and that the data protection aspect most commonly affected by the GDPR was informing the participants 
(changes reported by $84\%$ and $74\%$ of respondents in large companies and SMEs respectively). 
On the contrary, almost half of the participants in SMEs did not report any GDPR-related changes with regards to how the data is shared, how it is stored, and which controls are provided to the data subjects.

The analysis of the interviews revealed the following nuances in the state of GDPR compliance.

\paragraph{Rethinking data collection.}
As the GDPR has changed the data collection and management processes in  companies, it has pushed them to be more aware of their collection and retention policies around the different types of data they hold. A privacy expert stated: 

\begin{quote}
    ``We have been more aware of what kind of data we get from our customers. We're more aware of when to delete and for how long we need it. Also, we have become more aware of how much we actually need.''
\end{quote}

Since data collection requires management and compliance, as a result, many companies are avoiding to collect unnecessary data and are trying to erase it from their repositories by using automated or manual procedures. A senior manager remarked:

\begin{quote}
``Everything that we build from now on, we are aware of these things [GDPR compliance]. We have also put in procedures both manually and automatically that delete or anonymize data.''
\end{quote}

Furthermore, in line with the survey data, many interviewees from large companies mentioned that their companies provide control to the data subjects and make sure that they can ask for deletion from their systems. 

\begin{quote}
``They [data subjects] can ask us to get the data if they want to, we can transfer that out to them, if they want, we do not have any issues with that. Some people have written to us and ask us to erase all the data about them. And we do erase it, but we have some time scheduled to do it because we have some backup that also needs to go.''
\end{quote}

On the other hand, some interviewees mentioned that GDPR has not changed the data collection practices in their companies, 
either because in their perception, their companies did not store any personal data (e.g., being a business-to-business company), 
or they have outsourced the handling of their assets, including personal data, to third parties. 
A senior manager, when asked about the data collection changes in his company, articulated:

\begin{quote}
    ``We don't have anything, I mean, we don't carry any data.''
\end{quote}

\paragraph{Procedural complexities.}
During the interview, it became evident that many companies are facing different procedural challenges while handling the data as per the GDPR. A privacy expert articulated:

\begin{quote}
    ''[Data collection] has changed because now we must not have the data for a long time. And therefore we have to do a lot of things in a technical way to erase and all the things that we don't need anymore... 
    And therefore we have erased a lot of them and made some tools to eliminate all the unneeded data. So fewer data. Yes, but more specific and targeted to what we need in the business.''
\end{quote}

Since GDPR is stringent on the specificities of the data types that a company can hold, its retention period, data deletion after a certain period, and other controls exposed to the data subjects, 
some interviewees expressed the complexities it creates in their systems, making it challenging for them to implement.

\begin{quote}
	``[Regarding] the amount of data, we have changed a lot. We actually have a lot of de-validation or deletion tools that make sure that all this data is continuously deleted from our systems [...] 
	and also those tools are very difficult, well [we], make sure that they are up to date and keep working and that they don't delete anything they should not delete.''
\end{quote}



\paragraph{Guidance.}
During the interviews, some interviewees mentioned that their companies provide clear instructions to employees on GDPR compliance. They have portals where employees can read about the necessary measures to be taken when they are enabling GDPR in their work process. 

\begin{quote}
``We have a `Blackboard' [a portal] in the company. And then on [that] Blackboard, we have a specific area which is dealing with the GDPR and there are information papers for all the people and they can go in and see what do they need to look at and to know about.''
\end{quote}

Some companies also facilitate the opportunity to avail the consultation from a privacy expert in case the employee has doubts around the implementation process.  
\begin{quote}
    ``We have an internal ticket handling system where people can forward their GDPR related issues or security breaches, and other things to the GDPR group. And people use that to ask questions.''
\end{quote}

Still, the employees were expected to come forward and ask for clarifications themselves if they experienced problems. A developer articulated:
\begin{quote}
    ``From a product development point of view, there is no clear guidelines or no clear standard that our products can or cannot do this [...] I should say always it is the initiative of R\&D to go to legal and say, we have this idea of doing this or that, what should we be aware of?''
\end{quote}

%

At the same time, some interviewees talked about the lack of support for GDPR-related matters in their companies. One developer stated:

\begin{quote}
    ``I don't think we have a person responsible for security and privacy and GDPR and all the stuff that actually sits down and ensures that all this is in order''
\end{quote}

Some mentioned that there is a general lack of clear frameworks for the implementation of GDPR. It is difficult for their companies to comprehend and implement GDPR, as their companies are small in size and are operating without the support of a dedicated legal department. A senior manager:

\begin{quote}
    ``We felt that the guidelines were either very, very strict or very, very open to interpretation. So like everyone else, we were thinking, how do we approach this?''
\end{quote}





\paragraph{Burden on resources.}

It was evident in the interviews, that GDPR compliance requires a significant amount of time, money and expertise. 
A senior manager commented that smaller companies without such resources would not be able to properly ensure compliance:

\begin{quote}
    ``A number of our smaller competitors have it very difficult now because they find it very difficult to live up to the demands put upon them. 
    We are a little larger than many of them, and perhaps had a little more resources, both time, money, and intellectual resources as well to make sure we did comply.''
\end{quote}

Furthermore, some of the changes that were implemented for compliance reasons, such as the requirement to provide cookie notice on the company's website were perceived to be annoying and time-consuming rather than useful. 

\begin{quote}
    ``We have to wait a few weeks actually to get the texts ready and to make sure that the lawyers could see, what we were doing was correct. And all this kind of stuff annoys me a bit. ''
\end{quote}

\subsection{Effects of Pandemic}

\begin{figure}[htb]
	\centering
	    \centerline{\includegraphics[width=0.8\textwidth]{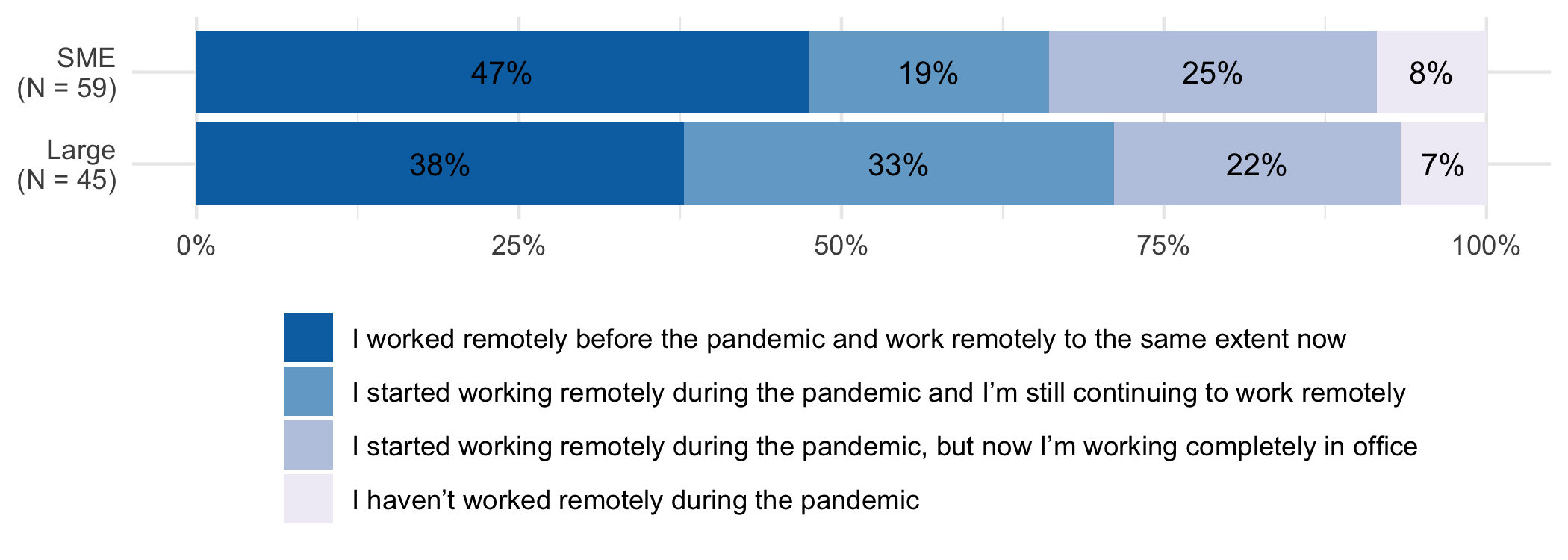}}
	    \caption{Remote work experience before and during the pandemic among the survey participants.\label{fig:pandemic_remote_experience}}
 \end{figure}

The survey results have shown that a vast majority of the participants had experience with remote work during the pandemic. As shown in \Cref{fig:pandemic_remote_experience}, a large part of them (38\% of respondents from large companies and 47\% from SMEs) were already working remotely even before the pandemic.

\begin{figure}[h]
	\centering
	    \centerline{\includegraphics[width=0.8\textwidth]{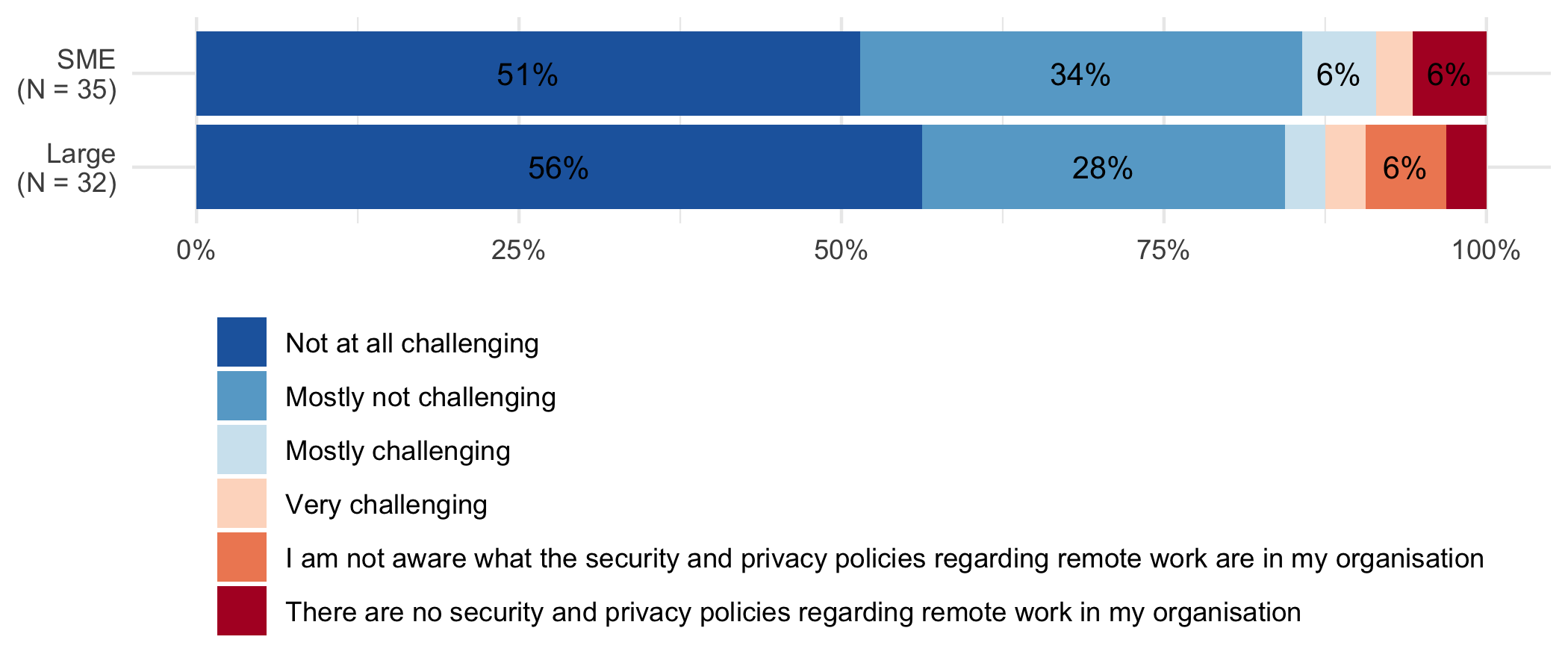}}
	    \caption{Experience with remote work policies among the survey participants.\label{fig:pandemic_remote_policies}}
 \end{figure}

The responses furthermore show that the majority of the respondents (85\% of all the participants who answered the question) did not experience issues with the security and privacy policies introduced for remote work, finding these policies either not challenging at all or mostly not challenging 
(\Cref{fig:pandemic_remote_policies}). 
 
\begin{figure}[htb]
	\centering
	    \centerline{\includegraphics[width=0.8\textwidth]{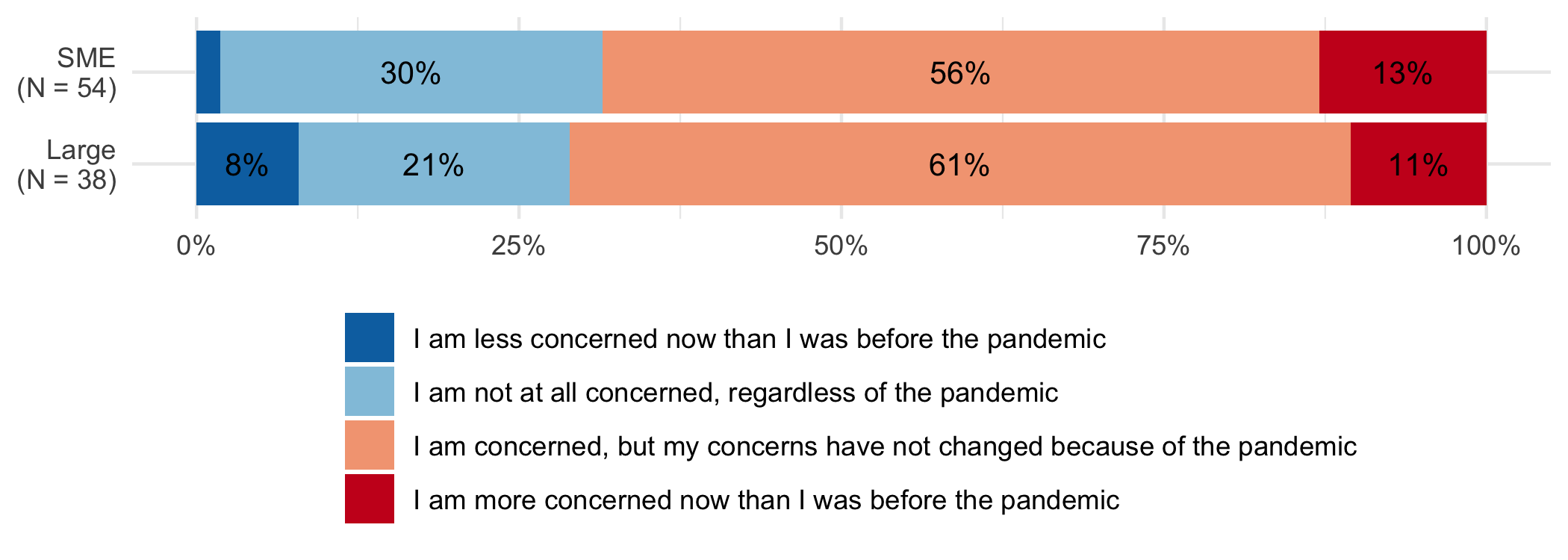}}
	    \caption{Pandemic-related security and privacy concerns among the survey participants.\label{fig:pandemic_concerns}}
 \end{figure}

Only a small percentage of respondents in both large companies (11\%) and SMEs (13\%) reported having increased concerns because of the pandemic and the remote work that followed (\Cref{fig:pandemic_concerns}).
 
The analysis of the interviews revealed the following nuances in the perspectives of the interviewees, on the influence of the pandemic.

\paragraph{Relying on pre-pandemic processes.} 

It was evident from our interviews that the pandemic has not changed the way of working, including security management, in many companies. A senior manager stated:

\begin{quote}
    ``We see that is okay to be working from home. We were also doing it before the pandemic. So for us, it was just a matter of scaling up for some of our systems to be able to accommodate the number of people working from home that was increasing dramatically.''
\end{quote}

Some interviewees mentioned that the infrastructure in their company had already been accustomed to remote work prior to the pandemic, e.g. by leveraging cloud and SaaS solutions, and they feel confident that the security measures are taken care of by the cloud and SaaS providers.

\begin{quote}
    ``Everything is cloud even the most sensitive part of the business, because then you know, that you have professionals handling those things. 
    [...] I don't even have to be on a VPN tunnel or something like that. I can work just on my computer.''
\end{quote}

\paragraph{Increased security.}

A few interviewees mention that their companies have implemented some changes to their security policies, such as mandating VPN usage when accessing the company data or forbidding the use of personal devices for work purposes. A security expert commented: 

\begin{quote}
    ``I think our biggest line of defense is the VPN planning [which] includes a firewall. So the corporate PC on the local network is isolated from the local event and you can only connect by the VPN.''
\end{quote}

A security expert commented on the pandemic's role in bringing awareness to cybersecurity in general, mentioning that it served as a catalyst for implementing some of the security controls that would otherwise take more time.

\begin{quote}
    ``The pandemic has accelerated our deployment of some security technologies that might otherwise have taken longer to deploy.
    So I think it's actually, in some ways benefited us and that we've now got some more improved remote access solutions and an improved level of security for remote workers than we would have done otherwise.''
\end{quote}

\paragraph{Working from home is more secure.} 

Another perspective that arose from interviews was the notion that working from home had in some aspects fewer security issues, due to limited physical access to company devices and sensitive documents.

\begin{quote}
    ``I would say I'm more afraid [when] I'm actually working at the office and leaving my laptop there. When the cleaning personnel comes in or some other guests are in [there] they could physically do something to my laptop. I think that's a bigger risk to manage [than] leaving the laptop at home.''
\end{quote}

\paragraph{Trust.} 

Several interviewees mentioned that since remote work has been an integral part of their company before the pandemic, there has been an inherent trust in the employees for not misusing the company devices or documents. A developer remarked, once again hinting at the role of trust in the Danish society:

\begin{quote}
    ``In Denmark it's like, we trust people to take their laptop home and we don't expect them to take any company data and stealing, of course [...] it's not like anyone is keeping eyes on you or something like that.''
\end{quote}


\section{Discussion}
\label{sec:discussion}
In this section, we reflect on our findings, identifying recurrent nuances throughout the different themes investigated in our studies, as well as discuss the limitations of our work.

\subsection{Recurrent nuances}

As presented in \Cref{sec:results},
particularly prominent was the mention of (i) \emph{trust}, which could also be perceived as a cultural characteristic of the demographics in our sample, and the emerging nuances of (ii) \emph{lack of awareness} of security and privacy risks, as well as (iii) \emph{lack of knowledge} about the appropriate countermeasures. We elaborate on these below.

\subsubsection{Trust}

Trust appeared as a recurring nuance in many of the themes after analyzing the data from the ethnographic interviews. When it comes to enabling security and privacy in the development cycle, senior managers in many organizations often choose to ``trust'' the developers adopting a ``let developers handle it'' mindset, assuming that the developers know all that needs to be done. Another occurrence of trust as an emerging nuance was apparent while evaluating the GDPR influence. Senior managers trust the developers in their organizations to be compliant with the GDPR while developing a new product or working on versioning an existing one. They feel that in case the developers find it challenging, they will come forward and ask for help---an approach that becomes a challenge if the developers themselves do not recognize their need for support, do not know whom exactly they could ask for such support or are otherwise reluctant to admit that they are struggling. Similarly, trust appeared as an integral part of the pandemic theme, where it was evident that Danish organizations trust their employees with their network security at home and feel assured that the employees are taking good care of the company's data and the physical devices. 

As the findings from the interviews suggest, such prevalence of trust, in terms of social cohesion (cf. \Cref{sec:introduction}), might be a reflection of a broader cultural trait of the Danish society (and possibly, more broadly, of Nordic countries), where people in the society are more likely to trust
each other, including their superiors or employees at their workplace~\cite{L13,sonderskov2014danish}. Such trust can have positive implications on the application of security and privacy measures. For example, employees' trust in the security team reduces the risk of not accepting or fighting against security policies and regulations---an issue that has been identified in previous studies in other cultural contexts~\cite{ashenden2013cisos,haney2018}. Furthermore, in cases where employees tend to trust their superiors and colleagues well enough to ask for help or report problems without fear of repercussions, issues in security and privacy measures are more likely to be identified and remedied promptly, provided that such transparent communication takes place. At the same time, trust that is misplaced can hurt security, especially without sufficient controls and accountability measures. Such an approach to security and privacy could lead to the compartmentalization of the implementation of security and privacy measures, potentially leading to a situation in which everyone feels that security and privacy measures are implemented regularly, whereas, in reality, it is left to developers to incorporate as they see fit. 

\subsubsection{Lack of awareness}

Lack of awareness and concern about security and privacy risks has manifested in different contexts in our study. Most notably, our investigation has shown that the pandemic did not lead to a significant increase in concerns or changes in security and privacy-related workflows, despite experts, both in Denmark and internationally, claiming an increased level of cyber-attacks and privacy issues~\cite{cfcs2020covid,interpol2020covid}.
Many of the participants in both interviews and surveys showed no concerns with security, saying that it creates no additional risks for them, as compared to the situation before the pandemic. For some interviewees, usage of cloud and cloud-based solutions was enough to ensure security for their landscape.

The lack of awareness has furthermore manifested in the discussion of changes to data protection policies required for GDPR compliance. As such, especially on the senior management level, many were not aware of what can potentially constitute personal data, and therefore subject to the GDPR. They were oblivious to the risk of non-compliance, which in turn creates a self-feeding circle in which the lack of awareness gets perpetuated across senior managers and developers.

The insufficient awareness or concern could easily correlate with the under-prioritization of security and privacy measures in the company, which has also shown by our studies (e.g., no defined budget for security, prioritize feature development over security). As such, the participants mentioned the perception of security needs as a distraction from the ``real'' business requirements and feeling that the senior management does not have it on their agenda, hindering regular updates of cybersecurity measures and eventually sacrificing security and privacy to cost optimization over the long period.

\subsubsection{Lack of knowledge}

Lack of knowledge on the appropriate security and privacy measures appeared as a ubiquitous nuance across multiple themes. 
As such, the developers commented on the unavailability of training to enhance their security and privacy skill-set, which is even more troubling in case the senior management feels that it is the responsibility of the developers to handle the security and privacy tasks on their own. Such a lack of necessary knowledge was also mentioned with regards to the GDPR compliance, with employees in many companies not being familiar with the procedural requirements to receive clarification on GDPR.
Similarly, the existing GDPR guidelines were perceived to be too vague, highlighting once again the need for better guidance.

While providing more security education measures, including training, might seem like a solution, there are also significant challenges with ensuring the effectiveness of such measures, such as their known problems of failing to engage the participants or provide them with knowledge and skills they can successfully apply outside of the training context~\cite{bada2019cyber}.  Indeed, as also shown by the results of our survey, a large share of participants did not find the training they attended to be useful. Furthermore, even if good training were available, in absence of clear management support and prioritization of security, the developers would have no incentive or desire to attend the training nor to acquire the needed knowledge otherwise.

\subsection{Limitations}
During our study, a sample was created to represent different sectors and sizes of organizations, and cover different participant' roles. Although the sample includes a range of participant roles, sectors, and size of organizations, the insights derived from the coding might not necessarily generalize to any of those variables. In particular, while we aimed for gender balance and reached out to other possible gender participants during the interviews, only male participants gave us their time-wise availability to conduct the interviews during the two months reserved for that.
Including more diverse perspectives on security and privacy would therefore be an important direction of future work.

Another challenge we experienced was the pandemic restrictions, which forced all the interviews to be conducted over video via MS-Teams rather than on-premise with the interviewees. 
The restriction over travel and in-person interactions might have affected the outcome of the interviews (e.g., missing non-verbal communication clues due to the shortcoming of online interviews versus face-to-face ones).


\section{Conclusions}
\label{sec:recommendations}
We conclude this paper by summarizing the main challenges that emerged from our
study. In particular, we deem these challenges likely to play key roles in the
social and economic norms of the more and more digitalized societies that will
emerge from the aftermaths of the pandemic.

\emph{Accounting for change} is the first and overarching topic of our investigation. Our results show that there is a need to develop guidelines and roadmaps that are not just designed to tackle a particular issue such as new legislation or the most recent crisis, but also are adaptable enough to be applied continuously to account for a variety of future changes. These roadmaps, for example, could result in guidelines for the companies in shaping their security training, ensuring regular updates and adaptations of their contents, as well as ongoing two-way collaborations between companies, researchers and public institutions.
Specifically, in the context of software development, such an accounting for change could be facilitated by methodologies such as Dev(Sec)Ops~\cite{devopssec,DevOps} and Site Reliability Engineering~\cite{SRE}, that already embrace change and security in their core process. While the interest in this kind of methodologies~\cite{Accelerate} has been increasing, more studies will be needed to understand how change could be integrated, especially for SMEs and for the more general picture in the digitalized society.

When investigating the adoption of the GDPR, we found that companies adopted a patchwork approach for handling the implementation of compliance measures to a sufficient extent, but many are still struggling with its adoption.
A more \emph{structured approach towards new regulations} is therefore needed for the forthcoming implementation of standards and regulations, 
e.g., via a creation of a task force constituted by the relevant stakeholders and lightweight conformity-assessment methods for basic security assurance~\cite{enisa2019software}.

%

Our results furthermore confirm and corroborate existing and well-know challenges like \emph{raising competences}. As previous research shows, awareness, while being an important first step towards improving security and privacy, is not sufficient, unless people are both provided with skills to cope with threats and are confident that they are capable of applying them~\cite{enisa2019cybersecurity}.
Our study shows a need for accessible training for developers and managers alike. To make the training relevant for the attendees, the security education measures should be tailored towards specific contexts, taking into account the general background and the needs of the developers that are about to participate, also ensuring that the participants would be able to easily translate the contents of the training into their daily tasks.
While the offer of test labs, cyber-ranges, documentation, and best practices has increased in the last two years, both on-premises and with cloud offerings~\cite{enisa2020etl},
particular attention must be given to check their effectiveness for training staff, simulating attacks, and testing multiple defense strategies.

%


Another aspect, emphasized also by previous research \cite{enisa2019cybersecurity} 
is the need of the \emph{managerial involvement}. In our study, we have witnessed that security and privacy measures are often perceived as a cost and therefore not properly prioritized. For the establishment of a proper security culture in the company, the involvement of management in the security decisions should be increased, ideally with senior management leading the company's security and privacy measures by their example and establishing a dedicated budget for security. While not all managers are expected to become security and privacy experts, they should have a basic awareness of security risks to drive the prioritization of security. They should also make sure that the developers feel incentivized to both implement the security measures they know of and also to improve their competences, e.g. by attending training, participating in conferences, and other educational opportunities.

Based on the mismatch between the perception of responsibilities with regards to security and privacy tasks we witnessed, we would recommend to managers also to foster as much as possible a \emph{transparent communication}. The expectations of both management and developers with regards to security and privacy responsibilities should be clearly communicated and agreed upon. Furthermore, efficient communication should be ensured for people seeking support with security and privacy-related task, so that they know whom they should turn to, be it \emph{security champions} \cite{thomas2018security} in their teams or a specifically assigned person of contact that handles security and privacy issues in the company.


Finally, we would like to conclude by emphasizing the \emph{role of culture} in security and privacy. Our study shows an effect of cultural contexts, such as the prevalence of trust in the companies towards external partners or employees, as a reflection of the importance of trust in general in Danish society. Further research into ways to support companies in their security and privacy practices while considering these cultural influences, including future studies with cross-cultural comparisons, might provide interesting insights. 

\bibliography{biblio}

\appendix

\section{Survey Questionnaire}
\label{sec:survey_questionnaire}
We present the questions of the survey. For space reasons, we are reporting only those questions in this questionnaire that are directly relevant to the research questions presented in this paper. The full questionnaire is available online (reference omitted for double-blind reviewing).

\paragraph{Survey questions}
\begin{enumerate}
    \item  What are your responsibilities in your organization?
    \emph{all, multiple choice, mandatory}
    \begin{itemize}
        \item  IT security related tasks
        \item Privacy/data protection related tasks
        \item Software development related tasks
        \item Management related tasks
        \item IT administrator related tasks
        \item Others, please specify 
    \end{itemize}

    \item Which industry/ sector is your organization in? 
    \emph{all, multiple selection, optional}
    \begin{itemize}
        \item Media \& Publishing
        \item Health care
        \item Financial services
        \item Software development
        \item Entertainment \& Music
        \item Education
        \item Manufacturing
        \item Consultancy
        \item Other, please specify
    \end{itemize}
    
    \item  Do you have a yearly budget allocated for Security \& Privacy needs?
    \emph{all, single choice}
    \begin{itemize}
        \item Yes 
        \item No
    \end{itemize}
    
    \item If YES, what percentage of your IT budget does it constitute?
    \emph{all, single choice}
    \begin{itemize}
        \item less than 1\% 
        \item 1\%-3\%
        \item 3\%-5\%
        \item More than 5\%
        \item We don’t have a defined security budget
        \item Not sure

    \end{itemize}

    \item How do you measure your cyber-security and privacy readiness? 
    \emph{Management, Sec, Priv, multiple choice, optional}
    \begin{itemize}
        \item We rely on the IT solutions derived from established security and privacy standards
        \item Internal method/framework/procedure
        \item We do not have any measure
    \end{itemize}
    
    \item If you rely on established standards to measure your cyber-security and privacy readiness, which ones do you use?
    \emph{Sec, Priv, multiple choice, optional}
    \begin{itemize}
        \item ISO/IEC 27001
        \item ISO 27701
        \item Center for Internet Security - Critical Security Controls (CIS CSC)
        \item Control Objectives for Information and Related Technologies (COBIT)
        \item Security for Industrial Automation and Control Systems (ANSI/ISA-62443)
        \item NIST Special Publication 800-53 (NIST SP 800-53)
        \item Payment Card Industry Data Security Standard (PCI DSS)
        \item UK National Cyber Security Centre (NCSC) 10 Steps
        \item UK National Health System (NHS) Digital Data Security and Protection Toolkit
        \item Cyber Assessment Framework (CAF)
        \item Information Assurance Small and Medium Enterprises (IASME)
        \item Host-Based Security System (HBSS)
        \item Structured Threat Information Expressions (STIX)
        \item Assured Compliance Accreditation Solutions (ACAS)
        \item Cyber Federated Model (CFM)
        \item Other, please specify

    \end{itemize}

    \item Has anything in the security and privacy practices of your organization changed since the introduction of the GDPR regarding the following aspects? \emph{Each of the subsections have \emph{yes, no \& not sure} options.}
    
    \emph{SecPriv, optional}
    \begin{itemize}
        \item Which data is collected by the organization
        \item How the data subjects are informed about data collection
        \item How the collected data is stored
        \item How the collected data is shared
        \item How the collected data is deleted
        \item Which controls are provided to the data subject
    \end{itemize}
    
    \item What is your experience with security and privacy awareness training at your company
    \emph{all, single choice, optional}
    \begin{itemize}
        \item I participated in training and found them useful
        \item I participated in training and did not find them useful
        \item I heard about available training but did not participate in them
        \item I am not aware of any security and privacy training at my organization
        \item Prefer not to answer
    \end{itemize}
    
     \item  When do you integrate security/privacy into your development practices? 
    \emph{Sec, Dev, single choice, optional}
    \begin{itemize}
        \item Early, from the initial phases
        \item Continuously during the development cycle
        \item After the fact
        \item Not at all
        \item Other, please specify
    \end{itemize}
    
    \item Are the methods/practices /standards for security and privacy protection in the development processes, always followed in all the situations? 
    \emph{Developer, Not Sec, Not priv, optional, single choice}
    \begin{itemize}
        \item Yes
        \item No
        \item Not sure
    \end{itemize}
    \item If NO, why?
    \emph{Developer, Not Sec, Not priv, optional, single choice}
    \begin{itemize}
        \item They are not always compatible with the functional requirements of our products
        \item I don't believe that they are helpful in protecting security and privacy
        \item They interfere with other workflows of my tasks and responsibilities
        \item They are too complicated to follow exactly as defined
        \item We don't have time or resources to follow them exactly as defined
        \item The management does not think they should be followed exactly as defined
        \item Other, please specify
    \end{itemize}

    \item To what extent has pandemic affected your working style, in particular remote working?
    \emph{all, single choice, optional}
    \begin{itemize}

        \item I worked remotely before the pandemic and work remotely to the same extent now
        \item I started working remotely during the pandemic, but now I’m working completely in office 
        \item I started working remotely during the pandemic and I’m still continuing to work remotely  
        \item I haven’t worked remotely during the pandemic 
      
    \end{itemize}
    
    \item If you started working remotely during the pandemic, how challenging do you find it to comply to the security and privacy policies of your organization regarding remote work?
    \emph{all NOT Sec \& NOT Priv, single choice, optional}
    \begin{itemize}
        \item Not at all challenging
        \item Mostly not challenging
        \item Mostly challenging
        \item Very challenging
        \item There are no security and privacy policies regarding remote work in my organization
        \item I am not aware of the security and privacy policies regarding remote work in my organization
    \end{itemize}
    
    \item  How did your concerns regarding security and privacy in your organizations change because of the pandemic?
    \emph{all NOT Sec \& NOT Priv, single choice, optional}
    \begin{itemize}
        \item I am more concerned now than I was before the pandemic
        \item I am concerned, but my concerns have not changed because of the pandemic
        \item I am less concerned now than I was before the pandemic
        \item I am not at all concerned, regardless of the pandemic
    \end{itemize}

\end{enumerate}

\section{Interview Guide}
\label{sec:interview_guide}
In this section, we present the Interview Guide that we used for the ethnographic interviews in our study.

\paragraph{Opening questions}

\emph{Purpose: To make the participant comfortable with the situation and present him/herself.}

\begin{enumerate}
    \item To begin with, please start by telling me about yourself? 
    \begin{itemize}
        \item What’s your role
        \item Day-to-Day life at work
        \item How long have you been working in this role and organization?
    \end{itemize}

\item Do you think about security and privacy in your regular work?
\begin{itemize}
    \item Is yes: How often and for what?
    \item If no: Why not?  
\end{itemize}

\item What is most important to you when it comes to security and privacy? (e.g. specific aspects of technology, business needs and constraints, etc.) 

\item How do you incorporate security and privacy in your daily work life?

\item Looking 2--3 years into the future, how would you expect the security and privacy needs to evolve?

\end{enumerate}

\paragraph{Questions to explore perceptions, motivation, perceived responsibility (outsourcing), and stress}

\emph{Purpose: To encourage the participant to share his/her perspective and realities of motivation and stress related to security usage in work life. Also, understanding the perceived responsibility of security and privacy when it is outsourced.}

\begin{enumerate}
    \item Do people in your organization feel that security and privacy are important?
    \item Do you feel that there are risks that are mitigated by security and privacy?
    \item How are security and privacy practices handled in your organization? (Outsourcing/SaaS/Insurance) 
    \item Do people face challenges when it comes to security and privacy(Implementing/ daily usage/ operations/ etc.) 
    \item Are there enough measures taken by your company to mitigate security and privacy problems? \begin{itemize}
        \item If yes: please describe
    	\item If no: Why not?
    \end{itemize}
	
    \item What do you think can be done to improve the situation?

\end{enumerate}

\paragraph{Questions to explore low adoption of continuous Sec adoption in the Development cycle}
\emph{Purpose: To understand the factors that inhibit the developers to do continuous integration of security in development cycles.}
\begin{enumerate}
    \item When do you typically implement security and privacy in your development cycle? 

    \item In your organization, do you have clear guidelines on precisely what methods and processes should be adopted to incorporate security in the development cycle? Could you broadly describe the methods and processes?

    \item Do you personally believe that your development cycles are including security considerations to the extent that they should?

    \item Do you get sufficient resources and training to enable this? 

    \item Do you follow Agile methodology in your development cycle? If so, do you think that Agile processes inhibit the developers to do continuous integration of security in development cycles? 

    \item Is management buy-in a contributing factor for this inhibition? According to you what role can management play in encouraging the adoption of security and privacy inclusion in development cycles?
\end{enumerate}

\paragraph{Questions to explore on GDPR influence}
\emph{Purpose: To encourage the participant to share their perspective on the influence of GDPR on the data collection.}
\begin{enumerate}
    \item  How has data collection behavior in your organization changed after GDPR?
    \begin{itemize}
        \item What all has changed? Could you provide some examples?
    \end{itemize}
	
    \item When is the data-collection management planning done? Is it done early in the stage with proactive planning or post fact ‘fixing’?
	
    \item How do people in your organization feel about giving controls to data-subjects?
    \begin{itemize}
        \item Do you have a well-defined approach for providing controls to data-subject? 
    \end{itemize}
\end{enumerate}

\paragraph{Questions to explore the impact of Pandemic} 
\emph{Purpose: To understand the security concerns of the participants due to higher remote working in Pandemic}
\begin{enumerate}
    \item How are people working in your organization? Are they more often in the office or prefer to work remotely?

    \item How do people feel about security adoption in your organization, while they are working more remotely?

    \item Has your organization crafted security policies to cover such a large-scale and sustained remote working condition?

    \item What all changes have been done in the security setup by your organization? Can you describe some of them?
\end{enumerate}

\end{document}